\RequirePackage{fix-cm}
\documentclass[reqno]{amsproc}
\usepackage{fixltx2e}
\numberwithin{equation}{section}

\usepackage{pstricks,pst-node}
\psset{unit=3ex}
\newcommand{\arr}{\lput{:U}{\begin{pspicture}(0,0)
    \psline(0,0.15)(.2,0)(0,-.15) \end{pspicture}}}
\newpsobject{ed}{pcline}{}
\newpsobject{th}{psline}{linewidth=.15}

\begin{document}
\title{The Arctic Circle Revisited}

\author{F. Colomo}
\address{I.N.F.N., Sezione di Firenze
and Dipartimento di Fisica, Universit\`a di Firenze\\
Via G. Sansone 1, 50019 Sesto Fiorentino (FI), Italy}
\email{colomo@fi.infn.it}

\author{A.G. Pronko}
\address{Saint Petersburg Department of Steklov Mathematical Institute
of Russian Academy of Sciences\\
Fontanka 27, 191023 Saint Petersburg, Russia}
\email{agp@pdmi.ras.ru}


\begin{abstract}
The problem of limit shapes in the six-vertex model with
domain wall boundary conditions is addressed by considering
a specially tailored bulk correlation function, the
emptiness formation probability.
A closed expression of this correlation function is given, both in terms
of certain determinant and multiple integral, which allows for
a systematic treatment of the limit shapes of the model for full
range of values of vertex weights. Specifically, we show that
for  vertex weights corresponding to the free-fermion line
on the phase diagram, the emptiness formation probability
is related  to a one-matrix model with a
triple logarithmic singularity, or Triple Penner model.
The saddle-point analysis of this model leads to the Arctic
Circle Theorem, and its generalization to the
Arctic Ellipses, known previously from domino tilings.
\end{abstract}

\maketitle

\section{Introduction}

The Arctic Circle has first appeared in the study of domino tilings
of large Aztec diamonds \cite{EKLP,JPS}. The name originates
from the fact that in most  configurations the dominoes are
`frozen' outside the circle inscribed into the diamond, while
the interior of  the circle
is a disordered, or `temperate',  zone. Further
investigations of the domino tilings of Aztec diamonds, such as details
of statistics near the circle, can be found in \cite{CEP,J1,J2}.
Here we mention that the Arctic Circle is a particular example of a
limit shape in dimer models, in the sense that it describes the shape
of a spatial phase separation of order and disorder. Apart from domino
tilings, many more examples
have been discussed recently, see, among others, papers
\cite{CKP,CLP,KO,KOS,OR}.

As long as only dimer models are considered, this amounts
to restrict to discrete free-fermionic models, although with nontrivial
boundary conditions. Indeed, many of them can be viewed
as a six-vertex model at its Free Fermion point
(the correspondence being however usually  not bijective),
with suitably chosen fixed boundary conditions. In particular, this is
the case of domino tilings of Aztec diamonds \cite{EKLP}, and
the corresponding boundary conditions of the six-vertex model
are the so-called Domain Wall Boundary Conditions (DWBC).
Hence the  problem of limit shapes extends  to
the six-vertex model with generic weights, and with fixed boundary
conditions, among which the case of DWBC is the most interesting.

Historically, the six-vertex model with DWBC was first considered
in paper \cite{K}  within the framework of Quantum
Inverse Scattering Method \cite{KBI} to prove the Gaudin
hypothesis for norms of Bethe states. The model was
subsequently solved in paper \cite{I} where a determinant formula
for the partition function was given; see also \cite{ICK} for
a detailed exposition. Quite independently, the model was later found, under
certain restrictions on the vertex weights, to be deeply related with
enumerations of alternating sign matrices (see, e.g., \cite{Br} for a
review) and, as already mentioned,
to domino tilings of Aztec diamonds \cite{EKLP}.

Concerning the problem of limit shapes for the six-vertex model with DWBC,
as far as the Free Fermion point is considered, the relation with
domino tilings provided apparently an indirect proof of
the corresponding Arctic Circle.
The non-bijective nature of the correspondence between the
two models asked for more direct
results, purposely for the free-fermion six-vertex model, see \cite{Zi1,FS,KP}.
Out of the Free Fermion point, however, only very few analytical results
are available, such as exact expressions for boundary
one-point \cite{BPZ} and two-point \cite{FP,CP1}
correlation functions. The present knowledge on the subject
is based mainly on numerics \cite{E,SZ,AR}; some steps towards finding
the limit shapes of the model have been done recently in \cite{PR}.

In the present note we propose a rather direct
strategy to address the problem: after briefly
reviewing the six-vertex model with DWBC,
we define a bulk correlation function, the
Emptiness Formation Probability (EFP), which discriminates the
ordered and disordered phase regions. We give for this correlation
function two equivalent representations, in terms of a determinant
and of a multiple integral. The core derivation of EFP is heavily based
on the Quantum Inverse Scattering Method \cite{KBI}, along the lines
of papers \cite{BPZ,CP1}; it is out of the scope
of the present paper, corresponding details being  given in a separate
publication \cite{CP4}. Here our aim is to demonstrate
how the limit shapes for the considered model
can be extracted from EFP in a suitable scaling limit, by
making use of ideas and techniques of Random Matrix Models.

To be more specific, and to establish a contact with previous results, we
specialize here our further discussion to the case of free-fermion
six-vertex model. We show that the asymptotic analysis of multiple integral
formula for EFP in the scaling limit reduces to a saddle-point problem for a
one-matrix model with a triple logarithmic singularity, or triple Penner
model. We argue that the limit shape corresponds to condensation of all
saddle-point solutions to a single point. This allows us
to recover the known Arctic Circle and Ellipses.

As a comment to our approach, it is to be stressed
that it is directly tailored on the six-vertex model, rather than
domino tilings.
For this reason it is not restricted to the free-fermion models, even if,
of course, further significant efforts might be necessary, essentially from
the point of view of Random Matrix Model reformulation,
for application to  more general situations.
On the basis of our previous results in \cite{CP2}, however,
the  application of the method to the particular case of the
so-called Ice Point of the model should be  straightforward.
This would provide the limit shape of alternating sign matrices.

\section{The model}

\subsection{The six-vertex model}

\begin{figure}[t]
\begin{pspicture}(0,-.5)(17,6)
\ed(0,5)(1,5)\arr \ed(1,5)(2,5)\arr \ed(1,4)(1,5)\arr \ed(1,5)(1,6)\arr
\ed(5,5)(4,5)\arr \ed(4,5)(3,5)\arr \ed(4,6)(4,5)\arr \ed(4,5)(4,4)\arr
\ed(6,5)(7,5)\arr \ed(7,5)(8,5)\arr \ed(7,6)(7,5)\arr \ed(7,5)(7,4)\arr
\ed(11,5)(10,5)\arr \ed(10,5)(9,5)\arr \ed(10,4)(10,5)\arr \ed(10,5)(10,6)\arr
\ed(12,5)(13,5)\arr \ed(14,5)(13,5)\arr \ed(13,5)(13,4)\arr \ed(13,5)(13,6)\arr
\ed(16,5)(17,5)\arr \ed(16,5)(15,5)\arr \ed(16,6)(16,5)\arr \ed(16,4)(16,5)\arr
\psline(0,2)(2,2)\psline(1,1)(1,3)
\th(5,2)(3,2)\th(4,3)(4,1)
\psline(6,2)(8,2)\th(7,3)(7,1)
\th(11,2)(9,2) \psline(10,1)(10,3)
\psline(12,2)(13,2)(13,3)\th(14,2)(13,2)(13,1)
\psline(16,1)(16,2)(17,2)\th(16,3)(16,2)(15,2)
\rput[B](1,-.25){$w_1$}
\rput[B](4,-.25){$w_2$}
\rput[B](7,-.25){$w_3$}
\rput[B](10,-.25){$w_4$}
\rput[B](13,-.25){$w_5$}
\rput[B](16,-.25){$w_6$}
\end{pspicture}
\caption{The six allowed types of vertices in terms of arrows and
lines, and their Boltzmann weights.}
\label{vertices}
\end{figure}
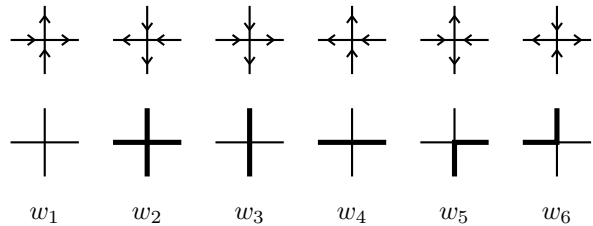

The six-vertex model  (for reviews, see \cite{LW,Ba})
is formulated on a square lattice
with arrows lying on edges, and obeying the so-called `ice-rule', namely,
the only admitted configurations are such that there are always two
arrows pointing away from, and two arrows pointing into, each lattice
vertex. An equivalent and  graphically simpler description
of the configurations of the model can be given
in terms of lines flowing through the vertices: for each arrow
pointing downward or to the left, draw a thick line on the
corresponding  edge. This line picture implements the `ice-rule'
in an automated way.
The six possible vertex states
and the Boltzmann weights $w_1,w_2,\dots,w_6$ assigned
to each vertex according
to its state are shown in Figure \ref{vertices}.

\subsection{Domain Wall Boundary Conditions}

\begin{figure}[t]
\begin{pspicture}(5,5)
\multirput(1,0)(1,0){4}{\ed(0,0)(0,1)\arr
\ed(0,5)(0,4)\arr}
\multirput(0,1)(0,1){4}{\ed(1,0)(0,0)\arr
\ed(4,0)(5,0)\arr}
\ed(2,4)(1,4)\arr \ed(2,4)(3,4)\arr \ed(3,4)(4,4)\arr
\ed(1,3)(2,3)\arr \ed(3,3)(2,3)\arr \ed(3,3)(4,3)\arr
\ed(2,2)(1,2)\arr \ed(3,2)(2,2)\arr \ed(4,2)(3,2)\arr
\ed(2,1)(1,1)\arr \ed(2,1)(3,1)\arr \ed(3,1)(4,1)\arr
\ed(1,4)(1,3)\arr \ed(1,2)(1,3)\arr \ed(1,1)(1,2)\arr
\ed(2,3)(2,4)\arr \ed(2,3)(2,2)\arr \ed(2,2)(2,1)\arr
\ed(3,4)(3,3)\arr \ed(3,2)(3,3)\arr \ed(3,1)(3,2)\arr
\ed(4,4)(4,3)\arr \ed(4,3)(4,2)\arr \ed(4,1)(4,2)\arr
\end{pspicture}
\hspace{3em}
\begin{pspicture}(5,5)
\multirput(1,0)(1,0){4}{\psline(0,0)(0,5)}
\multirput(0,1)(0,1){4}{\psline(0,0)(5,0)}
\th(1,5)(1,4)(0,4)
\th(2,5)(2,4)(1,4)(1,3)(0,3)
\th(3,5)(3,3)(2,3)(2,2)(0,2)
\th(4,5)(4,2)(2,2)(2,1)(0,1)
\end{pspicture}
\caption{A possible configuration of the six-vertex model with DWBC
at $N=4$, in terms of arrows and lines.}
\label{dwbcgrid}
\end{figure}
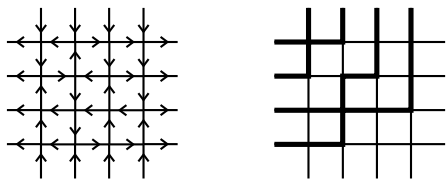

The Domain Wall Boundary Conditions (DWBC)
are imposed on the $N\times N$ square lattice by fixing
the  direction of  all arrows on the boundaries in a specific way.
Namely, the vertical arrows on the top and bottom
of the lattice point inward, while the horizontal arrows on the left
and right sides point outward. Equivalently, a generic configuration
of the model with  DWBC  can
be depicted by $N$ lines flowing from the upper boundary to the left one.
A possible state of the model both in terms
of arrows and of lines is shown in Figure \ref{dwbcgrid}.

\subsection{Partition function}

The partition function is defined, as usual, as a
sum over all possible arrow configurations, compatible with
the imposed DWBC, each configuration being assigned its Boltzmann weight,
given as the product of all the corresponding vertex weights,
\[
Z_N=\sum_{\substack{\text{arrow configurations}\\
\text{with DWBC}}}^{}\
w_1^{n_1}w_2^{n_2}\dots w_6^{n_6}\,.
\]
Here $n_1,n_2,\dots,n_6$ denote the numbers of vertices with weights
$w_1,w_2,\dots,w_6$, respectively, in each arrow configuration
($n_1+n_2+\dots +n_6=N^2$).

\subsection{Anisotropy parameter and phases of the model}

The six-vertex model with DWBC can be considered,
with no loss of generality,
with its weights invariant under the simultaneous reversal
of all arrows,
\[
w_1=w_2=:a\,,\qquad
w_3=w_4=:b\,,\qquad
w_5=w_6=:c\,.
\]
Under different choices of Boltzmann weights
the six-vertex model exhibits different  behaviours,
according to the value of the parameter $\Delta$, defined as
\[
\Delta=\frac{a^2+b^2-c^2}{2ab}\,.
\]
It is well known that
there are three physical regions or phases for the six-vertex model:
the ferroelectric phase, $\Delta>1$;
the anti-ferroelectric phase, $\Delta<-1$;
the disordered phase,
$-1<\Delta<1$. Here we restrict ourselves to
the disordered phase, where the Boltzmann weights are conveniently
parameterized as
\begin{equation} \label{param}
a=\sin(\lambda+\eta)\,,\qquad
b=\sin(\lambda-\eta)\,,\qquad
c=\sin 2\eta\,.
\end{equation}
With this choice one has $\Delta=\cos 2\eta$. The parameter $\lambda$
is the so-called spectral parameter
and $\eta$ is the crossing parameter.
The physical requirement of positive Boltzmann weights, in the disordered
regime, restricts the values of the  crossing and  spectral parameters
to $0<\eta<\pi/2$ and $\eta<\lambda<\pi-\eta$.

The special case  $\eta=\pi/4$  (or  $\Delta=0$)
is related  to  free fermions on a lattice, and there is
a well-known correspondence with dimers and domino tilings.
In particular, at $\lambda=\pi/2$, the $\Delta=0$
six-vertex  model with DWBC  is  related to
the domino tilings of Aztec diamond.
For arbitrary $\lambda\in[\pi/4,3\pi/4]$, we shall refer to
the $\Delta=0$ case as the Free Fermion line.

The case   $\eta=\pi/6$ (i.e. $\Delta=1/2$) and $\lambda=\pi/2$, where
all  weights are equal, $a=b=c$, is known as the Ice Point;
all configurations are given the same weight. In this case
there is a one to one correspondence between configurations of
the model with DWBC and $N\times N$ alternating sign matrices.

\subsection{Phase separation and limit shapes}

The six-vertex model exhibits spatial separation of phases
for a wide choice of fixed boundary conditions, and,
in particular, in the case of DWBC.
Roughly speaking, the effect is related to the fact
that ordered configurations on the boundary can induce,
through the ice-rule, a macroscopic order
inside the lattice.

The notion of phase separation acquires a precise meaning in the
scaling limit, that is the thermodynamic/continuum limit,
performed by sending the number of sites $N$ to infinity and
the lattice spacing to zero,
while keeping the total size of the lattice fixed, e.g., to $1$.
On a finite lattice, several macroscopic regions may appear,
which in the scaling limit  are expected to be sharply separated
by some curves, the so-called Arctic curves.

For the six-vertex model with DWBC
the shape of the Artic curve, or limit shape, has been found
rigorously only on  the Free Fermion line,
and for the closely related
domino tilings of  Aztec diamond \cite{JPS,CEP,Zi1,FS,KP}.
For generic values of weights the limit shapes are not known,
but the whole  picture is strongly supported both
numerically \cite{E, SZ, AR} and analytically \cite{KZ,Zi2,BF,PR}.

\section{Emptiness Formation Probability}\label{emptiness}

\subsection{Definition}

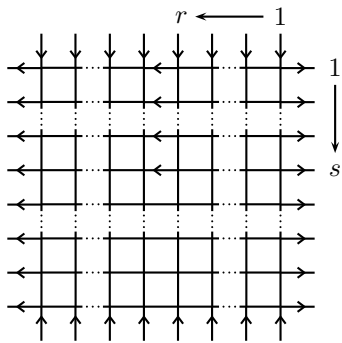
\begin{figure}[t]
\begin{pspicture}(10,10)
\multirput(1,0)(1,0){8}{\ed(0,0)(0,1)\arr \ed(0,9)(0,8)\arr}
\multirput(0,1)(0,1){8}{\ed(1,0)(0,0)\arr \ed(8,0)(9,0)\arr}
\multirput(0,8)(0,-1){4}{\ed(5,0)(4,0)\arr}
\multips(1,1)(1,0){8}{\psline(0,0)(0,2.2)}
\multips(1,1)(0,1){8}{\psline(0,0)(1.2,0)}
\multips(4,5)(0,1){4}{\psline(-1.2,0)(0,0)}
\multips(5,5)(0,1){4}{\psline(0,0)(1.2,0)}
\multips(3,1)(0,1){4}{\psline(-.2,0)(3.2,0)}
\multips(1,8)(1,0){8}{\psline(0,0)(0,-1.2)}
\multips(1,4)(1,0){8}{\psline(0,-.2)(0,2.2)}
\multips(8,1)(0,1){8}{\psline(0,0)(-1.2,0)}
\multips(1,3.3)(1,0){8}{\psline[linestyle=dotted,dotsep=.1](0,0)(0,.4)}
\multips(1,6.3)(1,0){8}{\psline[linestyle=dotted,dotsep=.1](0,0)(0,.4)}
\multips(2.3,1)(0,1){8}{\psline[linestyle=dotted,dotsep=.1](0,0)(.4,0)}
\multips(6.3,1)(0,1){8}{\psline[linestyle=dotted,dotsep=.1](0,0)(.4,0)}
\rput(5.1,9.5){$r$}
\rput(8,9.5){$1$}
\psline{->}(7.5,9.5)(5.5,9.5)
\rput(9.6,5){$s$}
\rput(9.6,8){$1$}
\psline{->}(9.6,7.5)(9.6,5.5)
\end{pspicture}
\caption{Emptiness Formation Probability. The sum in \eqref{FFPdef} is
performed over all configurations compatible with the drawn arrows.}
\label{ffpcorr}
\end{figure}

We shall use the following coordinates on the lattice: $r=1,\dots,N$
labels the vertical lines from right to left; $s=1,\dots,N$ labels
the horizontal lines from top to bottom.
We may now introduce the correlation function $F_N(r,s)$, measuring
the probability for the first $s$
horizontal edges between the $r$-th and $r+1$-th line to be all `full'
(i.e. thick in the line picture, or with a left arrow in the standard
picture of the six-vertex model):
\begin{equation}\label{FFPdef}
F_N(r,s) = \frac{1}{Z_N}\sum_{\substack{\text{`constrained'}\\ \text{
arrow configurations}\\ \text{with DWBC}}}^{}\
w_1^{n_1} w_2^{n_2}\dots w_6^{n_6}\,.
\end{equation}
Here the sum is performed  over all arrow configurations
on the $N\times N$ lattice, subjected to the restriction of DWBC, and
to the condition that all arrows on the first $s$ edges between
the $r$-th and $r+1$-th line should point left, see Figure \ref{ffpcorr}.

Although this correlation function may appear rather sophisticated,
it is computable in some closed form by means of the
Quantum Inverse Scattering Method, on which DWBC are indeed tailored.
It is the natural
adaptation of the Emptiness Formation Probability of quantum
spin chains to the present model. For this reason,
and to link to the common practice in the quantum integrable models
community, even if $F_N(r,s)$  actually describes `fullness'
formation probability, we shall call it Emptiness Formation Probability (EFP).

\subsection{Qualitative discussion  of $F_N(r,s)$}\label{discussion1}

Let us restrict ourselves to the  disordered regime, $-1<\Delta<1$,
for definiteness. From previous analytical and numerical work,
in the large $N$ limit the emergence of  a limit shape, in the form
of a continuous closed curve touching once each of the four sides of
the lattice, is expected. It follows that five regions emerge
in the lattice: a central region, enclosed by the curve, and four corner
regions, lying outside  the closed curve and delimited by the sides
of the lattice. The central region is disordered, while the four corners
are frozen, with mainly vertices of type 1, 3, 2, 4 (see Figure
\ref{vertices}) appearing
in the top-left, top-right, bottom-right and bottom-left corner,
respectively.

By construction, EFP is    expected to be almost one
in frozen  regions of type $1$, or $3$, bordering the top side of
the lattice, and to be rather  small  otherwise. DWBC
exclude a region of type $3$ to emerge in the upper part of the lattice.
Hence  $F_N(r,s)$  describes, at a given value of $r$,
 as $s$ increases,  a transition from a frozen region of vertices
of type $1$, where $F_N(r,s)\sim 1$,
to a generic region where $F_N(r,s)\sim 0$.

It follows that $F_N(r,s)$
can describe only the upper left portion of the closed curve,
between its top and  left contact points.
Nevertheless, it should be mentioned that the full curve can
be built from the knowledge of its top left portion,
just exploiting the crossing symmetry of the six-vertex model.
Hence  EFP, $F_N(r,s)$,  is well suited to describe  limit shapes.

\subsection{Some notations}

For a given choice of parameters $\lambda$, $\eta$
we  define
\[
\varphi := \frac{c}{ab} =
\frac{\sin2\eta}{\sin(\lambda+\eta)\sin(\lambda-\eta)}\,,
\]
and the integration measure
on the real line
\[
\mu(x) := \mathrm{e}^{x(\lambda-\pi/2)}
\frac{\sinh(\eta x)}{\sinh(\pi x/2)}\,,
\]
related to $\varphi$ as follows:
\[
\varphi = \int_{-\infty}^{\infty} \mu(x)\, dx\,.
\]
Let us introduce the complete set of monic orthogonal polynomial
$\left\{P_n(x)\right\}_{n=0,1,\dots}$  associated to the integration measure
$\mu(x)$, with the orthogonality relation
\[
\int_{-\infty}^{\infty} P_n(x) P_m(x) \mu(x)\, dx =
h_n \delta_{nm} \,.
\]
The square norms $h_n$ are completely determined by the measure
$\mu(x)$, and may be expressed, in principle,  in terms of its moments.
In the following we shall be interested in the complete
set of orthogonal polynomials $\left\{K_n(x)\right\}_{n=0,1,\dots}$
defined as
\[
K_n(x) = n!\, \varphi^{n+1} \frac{1}{h_n} P_n(x) \,.
\]
We moreover define
\[
\omega(\epsilon) :=
\frac{a}{b}\frac{\sin(\epsilon)}{\sin(\epsilon-2\eta)}\,,\qquad
{\tilde\omega}(\epsilon) :=
\frac{b}{a}\frac{\sin(\epsilon)}{\sin(\epsilon+2\eta)}\,.
\]
Note that the following relation holds
\begin{equation}\label{omegaomegatilde}
a^2\,\tilde{\omega}-2\Delta a b\,\tilde{\omega}\omega
+b^2\,\omega = 0\,,
\end{equation}
allowing to express $\tilde{\omega}$ in terms of $\omega$.

\subsection{Determinant representation}
For EFP in the six-vertex model with DWBC, the following representation holds:
\begin{multline}\label{FFPdet}
F_N(r,s) = (-1)^s
\det_{1\leq j,k \leq s}\left[K_{N-k}(\partial_{\epsilon_j})\right]
\, \prod_{j=1}^{s}
\frac{\left[\omega(\epsilon_j)\right]^{N-r}}{
\left[\omega(\epsilon_j)-1\right]^N}
\\ \times
\prod_{1\leq j < k\leq s}
\frac{\left[{\tilde \omega}(\epsilon_j)-1\right]
\left[\omega(\epsilon_k)-1
\right]}{{\tilde \omega}(\epsilon_j) \omega(\epsilon_k)-1}
\bigg|_{\epsilon_1=0,\dots,\epsilon_s=0}\,.
\end{multline}
This representation has been obtained in the framework
of the Quantum Inverse Scattering Method \cite{KBI},
along the lines of analogous
derivations worked out for one-point  and two-point boundary correlation
functions of the model \cite{BPZ,CP1}.
The details of the derivation can be found
in \cite{CP4}.

\subsection{The boundary correlation function}

If we consider  expression \eqref{FFPdet} when $s=1$, we recover the
boundary polarization, introduced and computed in \cite{BPZ}.
It is convenient to consider the closely related boundary correlation
function
\[
H_N(r):=F_N(r,1)-F_N(r-1,1)\,.
\]
As shown in \cite{BPZ,CP1}, the following representation
holds:
\[
H_N(r) = K_{N-1}(\partial_{\epsilon})
\frac{[\omega(\epsilon)]^{N-r}}{[\omega(\epsilon)-1]^{N-1} }
\bigg|_{\epsilon=0}\,.
\]
We define the corresponding generating function
\begin{equation}\label{generating}
h_N(z) := \sum_{r=1}^{N}\, H_N(r)\,z^{r-1}\,.
\end{equation}
Noticing that  $\omega(\epsilon)\to 0$ as $\epsilon \to 0$,
it can be shown that, given any arbitrary function
$f(z)$ regular in a neighbourhood of the origin, the following
inverse representation holds
\begin{equation}\label{integralrelation}
K_{N-1}(\partial_{\epsilon}) f(\omega(\epsilon))\bigg|_{\epsilon=0}
= \frac{1}{2\pi i}\oint_{C_0}
\frac{(z-1)^{N-1}}{z^N}\;  h_N(z)  f(z)\,dz\,.
\end{equation}
Here $C_0$ is a closed counterclockwise contour in the complex plane,
enclosing the origin, and no other singularity of the integrand.

\subsection{Multiple integral representation}

Plugging  \eqref{integralrelation} into representation
\eqref{FFPdet}, we readily obtain the following multiple integral
representation for EFP:
\begin{multline}\label{MIR}
F_N(r,s) = \left(-\frac{1}{2 \pi i}\right)^s
\oint_{C_0} \cdots \oint_{C_0} d^s\omega\,
\det_{1\leq j,k \leq s}
\left[h_{N-k+1}(\omega_j)\left(\frac{\omega_j-1}{\omega_j}\right)^{N-k}
\right]
\\ \times
\prod_{j=1}^s\frac{\omega_j^{N-r-1}}
{(\omega_j-1)^N}
\prod_{1\leq j<k\leq s} \frac{(\tilde{\omega}_j-1)(\omega_k-1)}
{\tilde{\omega}_j\omega_k-1}\,.
\end{multline}
Here $\tilde{\omega}_j$'s should be expressed in terms of $\omega_j$'s
through \eqref{omegaomegatilde}. Indeed, due to
\eqref{integralrelation}, relation \eqref{omegaomegatilde}
for functions $\omega(\epsilon)$, $\tilde{\omega}(\epsilon)$,
translates directly into the same  relation between   $\omega_j$ and
$\tilde{\omega}_j$, $j=1,\dots,s$.

Representation \eqref{MIR}, and all results in this
Section hold for any choice
of parameters $\lambda$ and $\eta$ within the disordered regime.
Moreover, by analytical continuation in parameters  $\lambda$ and $\eta$,
these results can be easily extended to all other regimes.

The determinant in expression \eqref{MIR} is a particular representation
of  the partition function of the six-vertex model with DWBC, when the
homogeneous limit is performed only on a subset of the spectral parameters
\cite{CP3}. The structure of the previous multiple integral
representation therefore closely recalls analogous ones for the
Heisenberg XXZ quantum spin chain correlation functions
\cite{JM,KMT}.

For generic values of $\lambda$ and $\eta$,
the orthogonal polynomials $K_n(x)$, or  the
generating function $h_N(z)$, are known only in terms of rather
implicit representations.
Fortunately, there are three
notable exceptions \cite{CP2}: the Free Fermion line
($\eta=\pi/4$, $-\pi/4<\lambda<\pi/4$, $\Delta=0$),
the Ice Point ($\eta=\pi/6$, $\lambda=\pi/2$, $\Delta=1/2$),
and the Dual Ice Point ($\eta=\pi/3$, $\lambda=\pi/2$, $\Delta=-1/2$).
In these three cases, the $K_n(x)$ turn out to be classical orthogonal
polynomials, namely  Meixner-Pollaczek, Continuous Hahn and Continuous
Dual Hahn polynomial, respectively.
Correspondingly, the generating function can be represented explicitly
in terms of terminating  hypergeometric functions that may
simplify considerably further evaluation of EFP.
In the next Section we shall focus on the case of Free Fermion line.

\section{Multiple integral representation at $\Delta=0$}

\subsection{Specialization to $\eta=\pi/4$}
We shall now restrict ourselves to  the case $\eta=\pi/4$.
We have $\Delta=0$, and the six-vertex model
reduces to a model of free fermions on the lattice. The parameter
$\lambda$ can still assume any value in the interval $(-\pi/4,\pi/4)$.
It is convenient to trade $\lambda$ for the new parameter
\[
\tau=\tan^2(\lambda-\pi/4)\,, \qquad 0<\tau<\infty\,.
\]
The symmetric point (related to the domino tiling of Aztec Diamond)
corresponds now to $\tau=1$.
For generic values of $\tau$ we have:
\[
{\tilde \omega} =-\tau\omega\,.
\]
The generating function \eqref{generating}
is known explicitely (see \cite{CP2} for details):
\[
h_N(z) = \left(\frac{1+\tau z}{1+\tau}\right)^{N-1}\,.
\]
Plugging this expression into \eqref{MIR}, we get
\begin{multline}\label{MIRforFF1}
F_N(r,s) = \left(-\frac{1}{2 \pi i}\right)^s
\oint_{C_0}\cdots\oint_{C_0} d^s\omega\,
\det_{1\leq j,k \leq s}\left\{
\left[\frac{(1+\tau\omega_j)(\omega_j-1)}{(1+\tau)\omega_j}\right]^{N-k}
\right\}
\\ \times
\prod_{j=1}^s\frac{\omega_j^{N-r-1}}
{(\omega_j-1)^N}
\prod_{1\leq j<k\leq s} \frac{(1+\tau\omega_j)(\omega_k-1)}
{1+\tau\omega_j\omega_k}\,.
\end{multline}

\subsection{Symmetrization}

After extracting a common factor
\[
\prod_{j=1}^s
\left[\frac{(1+\tau\omega_j)(\omega_j-1)}{(1+\tau)\omega_j}\right]^{N-s}
\]
from the determinant in \eqref{MIRforFF1}, we recognize it
to be of Vandermonde type.
We can therefore collect from the integrand of \eqref{MIRforFF1}
the double product
\[
\prod_{1\leq j<k\leq s} \left[
\frac{(1+\tau\omega_j)(\omega_j-1)}{(1+\tau)\omega_j}-
\frac{(1+\tau\omega_k)(\omega_k-1)}{(1+\tau)\omega_k}
\right]
\frac{(1+\tau\omega_j)(\omega_k-1)}{1+\tau\omega_j\omega_k}\,.
\]
Noticing that the integration and the remaining of integrand are fully
symmetric under permutation of variables $\omega_1,\dots,\omega_j$,
we can perform total symmetrization of the previous double product
over all its variables, with the result
\[
\frac{1}{s!} (-1)^{s(s-1)/2} \prod_{j=1}^s \frac{1}{\omega_j^{s-1}}
\prod_{1\leq j<k\leq s} (\omega_j-\omega_k)^2\,.
\]
Hence, we finally obtain the following representation for EFP
on the Free Fermion line:
\begin{multline}\label{MIRforFF2}
F_N(r,s) = \frac{(-1)^{s(s+1)/2}}{s! (1+\tau)^{s(N-s)}(2 \pi i)^s}
\\ \times
\oint_{C_0} \cdots\oint_{C_0} d^s\omega\,
\prod_{1\leq j<k\leq s} (\omega_j-\omega_k)^2
\prod_{j=1}^s
\frac{(1+\tau\omega_j)^{N-s}}{(\omega_j-1)^s\, \omega_j^r}\,.
\end{multline}
The appearance of a squared Vandermonde determinant in this
expression naturally
recalls the  partition functions of $s \times s$
Random Matrix Models.

\section{Triple Penner model and Arctic Ellipses}

\subsection{Scaling limit}\label{scaling}

We shall now address the asymptotic  behaviour of expression \eqref{MIRforFF2}
for EFP in the $\Delta=0$ case.
We are interested in the limit $N,r,s\to\infty$, while
keeping the ratios
\[
r/N=x\,, \qquad s/N=y\,,
\]
fixed. In this limit, $x,y\in[0,1]$ will parameterize
the unit square to which the lattice is rescaled.
Correspondingly EFP is expected
to approach a limit function
\[
F(x,y) := \lim_{N\to\infty}F_N(xN, yN)\,,\qquad x, y\in[0,1]\,.
\]
We shall exploit  the standard approach developed for instance
in the investigation of asymptotic behaviour for Random Matrix Models.
Before this let us however point out some facts which holds
already for any finite value of $s$.

\subsection{A useful identity}

Consider  the quantity
\begin{multline*}
I_N(r,s) := \frac{(-1)^{s(s+1)/2}}{s! (1+\tau)^{s(N-s)}(2 \pi i)^s}
\\ \times
\oint_{C_1} \cdots\oint_{C_1} d^s\omega
\prod_{1\leq j<k\leq s} (\omega_j-\omega_k)^2
\prod_{j=1}^s
\frac{(1+\tau\omega_j)^{N-s}}{(\omega_j-1)^s\, \omega_j^r}\,,
\end{multline*}
which differs from \eqref{MIRforFF2} only in the integration contours.
Here $C_1$ is a closed, \textit{clockwise} oriented contour
(note the change in orientation) in the complex plane enclosing point
$\omega=1$,  and no other singularity
of the integrand.
We have the identity
\begin{equation}\label{I=1}
I_N(r,s)=1
\end{equation}
for any integer $r,s=1,\dots,N$.
The simplest way to prove the previous identity is
by shifting $\omega_j\to\omega_j+1$, and rewriting $I_N(r,s)$
as  an Hankel determinant; indeed we have
\[
I_N(r,s) = \frac{(-1)^{s(s-1)/2}}{(1+\tau)^{s(N-s)}}
\det_{1\leq j,k \leq s}
\left[\frac{1}{2\pi i}\oint_{C_0}\frac{\omega^{j+k-2-s}(1+\tau+\tau\omega)^{N-s}}
{(1+\omega)^r}\,d\omega
\right]\,.
\]
The entries of the  Hankel matrix vanish for $j+k>s+1$, and hence
the determinant is simply given by the product of the antidiagonal entries,
$j+k=s+1$ (modulo a sign $(-1)^{s(s-1)/2}$ emerging from
the permutation of all columns). Identity \eqref{I=1} follows immediately.

\subsection{Saddle-point evaluation for large $N$ and finite $s$}
\label{discussion2}
 When using  the saddle-point method in
variables $\omega_1,\dots,\omega_s$  to evaluate the behaviour
of $F_N(r,s)$ for large $N$ and $r$, and finite $s$, it is rather easy
to see that  the saddle-point equations decouple at leading order,
and that each saddle-point will be on the real axis,
contributing with a factor $\mathrm{e}^{-N S_j}$ with  $S_j$ positive.

If a given saddle-point is smaller than $1$,
the contour $C_0$ can be deformed through the saddle-point without
encountering any pole, and its contribution  will vanish as
$\mathrm{e}^{-N S_j}$ in the large $N$
limit. If however the saddle-point, still on the real axis,
happens to be  larger than $1$, the deformation of the contour $C_0$
through the saddle-point will pick up the contribution of the pole
at $\omega=1$  (with a reversed orientation of the contour),
and the $j$-th integral will behave as $1+\mathrm{e}^{-N S_j}$.
Hence, in the large $N$ limit (at fixed $s$) the quantity
$F_N(r,s$) will vanish unless all the saddle-points are
greater than $1$, in which case $F_N(r,s)\sim I_N(r,s)=1$.
Note that in the present situation the $s$ saddle-points coincide.
A detailed analysis shows that in this case the position of the $s$
saddle-points depends on the value $x=r/N$ as $\omega_0=\frac{x}{\tau(1-x)}$.
In correspondence to the value $x_0=\frac{\tau}{1+\tau}$,
for which these saddle-points  are exactly $1$, the function
$F(x,0)$ has a step discontinuity.  More precisely, it is easy to show
that for $x\in[0,1]$, $F(x,0)=\theta(x-x_0)$, where $\theta(x)$ is
Heaviside step function. From a physical point of view $x_0$
is the contact point between the limit shape and the boundary.
What have been discussed here can easily be verified in the case $s=1$.
The extension to finite $s>1$ is rather direct as well.

\subsection{Saddle-point equation}

Having in mind the analogy with $s \times s$ Random Matrix Models,
and the scaling limit specified in Section \ref{scaling},
we rewrite our expression for $F_N(r,s)$ at $\Delta=0$ as follows:
\begin{multline}\label{MIRasRMM}
F_N(r,s) =
\frac{(-1)^{s(s+1)/2}}{s! (1+\tau)^{s^2(1/y-1)}(2 \pi i)^s}
\oint_{C_0} \cdots\oint_{C_0} d^s\omega\,
\exp\bigg\{\sum_{\substack{j,k=1\\ j\ne k}}^{s} \ln|\omega_j-\omega_k|
\\
+ s \sum_{j=1}^s \left[\left(\frac{1}{y}-1\right)\ln(\tau\omega_j+1)
-\ln(\omega_j-1)-\frac{x}{y}\ln(\omega_j)
\right]
\bigg\}\,.
\end{multline}
Both sums in the exponent are  $O(s^2)$.
The corresponding (coupled) saddle-point equations read
\begin{equation}\label{SPE}
\frac{1}{\omega_j-1}+
\frac{x/y}{\omega_j}-\frac{(1/y-1)\tau}{\tau\omega_j+1} =
\frac{2}{s}\sum_{\substack{k=1\\ k\ne j}}^{s}\frac{1}{\omega_j-\omega_k}\,.
\end{equation}

A standard physical picture reinterprets the saddle-point equations
as the equilibrium condition for the positions of
$s$ charged particle confined to the real axis, with logarithmic
electrostatic repulsion, in an external potential. In the present case
the latter can be seen as generated by three external
charges, $1$, $x/y$, and $-(1/y-1)$ at positions $1$, $0$, and $-1/\tau$,
respectively.
It is natural to refer to this model as the triple Penner
model. Although the simple Penner \cite{P} matrix model has been widely
investigated,
not so much is known about the much more complicate
double Penner model \cite{M,PW}.
We have not been able to trace any previous study concerning
the triple Penner model.

\subsection{The exact Green function at finite $s$}

To investigate the structure of solutions of the saddle-point equations
\eqref{SPE} for large $s$ we first introduce the Green function
\[
G_s(z) = \frac{1}{s}\sum_{j=1}^s \frac{1}{z-\omega_j}\,,
\]
which, if the $\omega_j$'s solves \eqref{SPE},  has to satisfy the
differential equation:
\begin{multline}\label{diffeqforG}
z(z-1)(\tau z+1)\left[ s G'_s(z)+s^2G_s^2(z)\right]
-s(\alpha z^2+\beta z+\gamma)s G_s(z)
\\
= \left[\tau s(s-1)-\alpha s^2\right]z
+(1-\tau)s(s-1)-\beta s^2
+\Omega\left[2\tau s (s-1)-\alpha s^2\right]\,.
\end{multline}
The coefficients $\alpha$, $\beta$ and $\gamma$
are readily obtained as the coefficients of the second order polynomial
appearing in the numerator, when  setting to common denominator the left
hand side of \eqref{SPE}. We give them explicitly for later convenience:
\[
\alpha=\tau \left(2-\frac{1-x}{y}\right)\,, \qquad
\beta=\frac{\tau}{y}+(1-\tau)\left(1+\frac{x}{y}\right)\,, \qquad
\gamma=-\frac{x}{y}\,.
\]
The derivation of the differential equation is very standard
(see, e.g., \cite{SD}). The left hand side is built by suitably combining
the explicit definition of the Green function and its derivative.
The result has to be a polynomial of the first degree in $z$,
whose coefficients
are constructed by matching the leading and first subleading behaviour
of the left  hand side as $|z|\to\infty$.

\subsection{The first moment $\Omega$}

The quantity
${\Omega}$ appearing in \eqref{diffeqforG} is defined as the first moment
of the solutions of the saddle-point equations:
\[
{\Omega }:=\frac{1}{s}\sum_{j=1}^{s}\omega_j\,.
\]
It is  related in a obvious way to the first subleading coefficient
of $G_s(z)$; indeed, from the definition of the Green function, it is
evident that
\[
G_s(z) =  \frac{1}{z}+ \Bigg(\frac{1}{s}\sum_{j=1}^s \omega_j\Bigg)
\frac{1}{z^2} +
O(z^{-3})\,,\qquad |z|\to\infty\,.
\]
It is worth to emphasize that ${\Omega }$ is still unknown,
and that in principle its value  should be determined  self consistently
by first working out the explicit solution of $G_s(z)$ (which will depend
implicitly on ${\Omega }$), from  \eqref{diffeqforG} and then demanding that
$\frac{1}{s}\sum_{j=1}^s \omega_j$ evaluated from this solution coincides
with $\Omega $.
The appearance of the undetermined parameter   ${\Omega }$
is a manifestation of the `two-cuts' nature of the Random Matrix Model
related to \eqref{MIRasRMM}, see, e.g., par. 6.7 of \cite{D1}.

\subsection{The asymptotic Green function}
We are now in condition to perform the large $s$ (and large $N$, $r$) limit
at fixed $x,y$.
In the limit, we can neglect terms of order $O(s)$ in the
differential equation \eqref{diffeqforG}, which therefore reduces
to an algebraic equation for  the limiting Green function $G(z)$:
\begin{equation}
z(z-1)(\tau z+1)[ G(z)]^2
-(\alpha z^2+\beta z+\gamma) G(z) = (\tau-\alpha)z
+(1-\tau-\beta)
+\Omega (2\tau -\alpha)\,.
\end{equation}
The previous algebraic equation has to be supplemented by
the normalization condition
\begin{equation}\label{normalization}
G(z)\sim\frac{1}{z}\,, \qquad |z|\to\infty\,.
\end{equation}
Hence the Green function describing the large $s$ asymptotic
distribution of solutions for the saddle equation \eqref{SPE} reads:
\begin{multline}\label{Greeneq1}
G(z) = \frac{1}{2z(z-1)(\tau z+1)}\Big\{
(\alpha z^2 +\beta z+\gamma)
\\
+\sqrt{(\alpha z^2 +\beta z+\gamma)^2
+4z(z-1)(\tau z+1)[(\tau-\alpha)z+
\Omega (2\tau-\alpha)+1-\tau-\beta]}
\Big\}.
\end{multline}
We have selected the positive branch of the square root,
to satisfy the normalization condition (note that the coefficient
of $z^4$ under the square root is $(\alpha-2\tau)^2$, and $\alpha-2\tau$
is negative for any $x, y\in[0,1]$).
However,  the expression for $G(z)$ is not completely specified yet,
because $\Omega $ is still undetermined.

\subsection{Limit shape and condensation of roots}\label{discussion3}

The polynomial under the square root is of fourth order, hence $G(z)$
will have in general two cuts in the complex plane. The emergence
of a two-cut problem was already expected from the appearance
of the undetermined first moment $\Omega $ in  \eqref{diffeqforG}.
The discontinuity
of $G(z)$ across these cuts defines, when positive, the density
of solutions of the saddle-point equations \eqref{SPE} when $s\to\infty$.
The problem of explicitly finding this density, for arbitrary
$\alpha$, $\beta$, $\gamma$ (or $x$, $y$), is a formidable one, not to
mention the evaluation of the corresponding `free energy', and of
the saddle-point contribution to the integral in \eqref{MIRasRMM}.
But our aim is much more modest, since we are presently interested only
in the expression of the limit
shape, i.e. in the curve in the square $x,y\in[0,1]$, delimiting regions
where $F(x,y)=0$ from regions where  $F(x,y)=1$.
Of course we are here somehow assuming that the transition of $F(x,y)$
from $0$ to $1$ is stepwise in the scaling limit, but this is supported
both by the physical interpretation of EFP
(in the disordered region,
by definition, the number of `thin' lines   is macroscopic, and the
probability of finding no `thin' horizontal edges immediately vanishes
in the scaling limit) and by the discussion of Section \ref{discussion2}.

As explained in the discussion of the double Penner model in paper
\cite{PW}, the logarithmic wells in the potential
can behave as condensation germs for the saddle-point solutions.
In our case, this can role can be played only by the `charge' at $\omega=1$
in the Penner potential since the charge at $\omega=-1/\tau$
is always repulsive, while the one at $\omega=0$ is larger than
$1$, at least in the region of interest.
\cite{PW} have shown that condensation can occur only
for charges less than or equal to $1$,
since this will be the fraction of condensed solutions.
This consideration, together with the expected stepwise behaviour
and the discussion in Section \ref{discussion2}, suggest the following picture
for the evolution of saddle-point solution density from the
disordered region, $F(x,y)\sim 0$,  to the upper
left frozen region, $F(x,y)\sim 1$:
in the disordered region there is a macroscopic fraction of solutions
which are real and smaller than $1$, while in the upper left frozen region
this fraction vanish. On the basis of the discussion here and in
Sections \ref{discussion1} and \ref{discussion2},
we shall assume that at the transition between the two regions
all  saddle-point solutions have  condensed at $\omega=1$.

\subsection{Main assumption}

We claim that  the Arctic curve in the
square $x,y\in[0,1]$ separating the disordered phase from
the upper left frozen  phase is defined by the condition
that all solutions of the saddle-point equation lies at $\omega=1$.

In the derivation of the limit shape, this is  indeed the only  assumption
to which we are unable to provide  a proof.
There is in fact no guarantee, at this level, for this possibility to
occur, and limit shapes could in principle emerge from a different
condition. But if for some values of  $x,y\in[0,1]$ we
have all solutions of the saddle-point equation condensing at $\omega=1$,
then this   provides a transition mechanism from
$0$ to $1$ for $F(x,y)$, and this might  correspondingly define some
limit shape.

If all saddle-point solutions condensate at $\omega=1$,
then we obviously have:
\[
\Omega =1\,.
\]
Moreover, the complicate expression \eqref{Greeneq1} for $G(z)$
should simply reduce  to
\begin{equation}\label{Greeneq2}
G(z)=\frac{1}{z-1}\,,
\end{equation}
since we expect to have no cuts, and only one pole at $z=1$ with
unit residue.

\subsection{Arctic Ellipses}
Consider the quartic polynomial under the square root in \eqref{Greeneq1}.
It is convenient to rewrite it in terms of
\begin{equation}\label{alphatilde}
\begin{split}
{\tilde{\alpha}} &:= 2\tau-\alpha = \tau\,\frac{1-x}{y}\,,\\
{\tilde{\beta}} &:= 2-\beta = \tau\,\frac{x+y-1}{y}+\frac{y-x}{y}\,,\\
{\tilde{\gamma}} &:= -\gamma = \frac{x}{y}\,.
\end{split}
\end{equation}
Note that ${\tilde{\alpha}}$ and ${\tilde{\gamma}}$ are always positive for
$x,y\in[0,1]$. When $\Omega =1$, our quartic polynomial reads
\[
{\tilde{\alpha}}^2 z^4+ 2 {\tilde{\alpha}} {\tilde{\beta}} z^3 +
({\tilde{\beta}}^2+2 {\tilde{\alpha}} {\tilde{\gamma}})z^2 +
2 {\tilde{\beta}} {\tilde{\gamma}} z +{\tilde{\gamma}}^2\,,
\]
which may be equivalently rewritten as
\[
({\tilde{\alpha}} z^2 +{\tilde{\beta}} z +{\tilde{\gamma}})^2 \,.
\]
We see that the quartic polynomial reduces to a perfect square,
and hence,  when $\Omega =1$,  the two cuts of $G(z)$ disappear,
as expected.

Now, when  $\Omega =1$, in our new notations,
the Green function reads:
\begin{equation}\label{Greeneq3}
G(z)=
\frac{[(2\tau-{\tilde{\alpha}}) z^2 + (2-{\tilde{\beta}}) z -
{\tilde{\gamma}}]+\sqrt{({\tilde{\alpha}} z^2 +{\tilde{\beta}} z
+{\tilde{\gamma}})^2 }}
{2 z(\tau z +1)(z-1)}\,.
\end{equation}
We now require the coefficients ${\tilde{\alpha}}$, ${\tilde{\beta}}$,
${\tilde{\gamma}}$ to be such that
the polynomial under the square root combines
with the first part of the numerator in \eqref{Greeneq3} to give
$2 z (\tau z+1)$ and simplify  the Green function according
to \eqref{Greeneq2}. Once we have chosen a given branch of the square root
(the positive one, in order to satisfy normalization condition
\eqref{normalization}),
it is obvious that the  required simplification can occur for any $z$
in the complex plane  only if
the second order polynomial ${\tilde{\alpha}} z^2+
{\tilde{\beta}} z +{\tilde{\gamma}}$ does not change its sign, i.e.
only if its two roots coincide, implying:
\[
{\tilde{\beta}}^2 - 4 {\tilde{\alpha}} {\tilde{\gamma}}=0\,.
\]
Rewriting the last relation in terms
of $x$, $y$, through \eqref{alphatilde}, we readily get
\[
(1+\tau)^2 x^2+(1+\tau)^2 y^2-2(1-\tau^2) x y-2\tau(1+\tau)x-2\tau(1+\tau)y
+\tau^2 = 0\,.
\]
We have therefore recovered the limit shape, which in this Free Fermion
case is the well-known Arctic Ellipse (Arctic Circle for $\tau=1$)
\cite{JPS,CEP}.
We recall that, as discussed in Section \ref{discussion1},
$F(x,y)$ is non-vanishing only in the upper
left region of the unit square. Therefore, concerning EFP,
only the upper left portion of the Arctic curve, between the
two contact points at  $(\frac{\tau}{1+\tau},0)$ and $(1,\frac{1}{1+\tau})$,
is relevant.

\section{Concluding remarks}

Our starting point has been the definition of a
relatively simple but relevant correlation function
for the six-vertex model with DWBC, the Emptiness Formation
Probability. We have provided
both a determinant representation and a multiple integral
representation for the proposed correlation function. This is the first
example in literature of a bulk (as opposed to boundary)
correlation function  for the considered model, for generic weights.

The multiple integral representation, specialized
to the Free Fermion case,  has been studied in the scaling limit.
In the standard picture of Random Matrix Models, we
recognize the emergence
of a triple Penner model. Assuming condensation of the roots of saddle
point equations in correspondence  to a limit shape, we  recover the
well-known
Arctic Circle and Ellipse. It would be interesting to investigate whether
universality considerations of Random Matrix Models (see, e.g., \cite{D2})
can be extended to the Penner model in the  neighbourhood of its logarithmic
singularities. This would imply directly the results of
\cite{CEP,J1,J2} on the Tracy-Widom distribution and
the Airy process, emerging in a suitably rescaled
neighbourhood of the Arctic Ellipse.

It is worth to stress that the multiple integral representation
for EFP presented in Section \ref{emptiness}
can be studied  beyond the usual Free Fermion situation.
We expect that condensation of roots
of the saddle point equation in correspondence of  the limit shape
is a general phenomenon. We believe that  this assumption could be
of importance  in addressing the problem of limit shapes in the
six-vertex model with DWBC.

Our derivation of the  limit shape in the Free Fermion case uses  the
explicit knowledge of function $h_N(z)$, standing  in the multiple
integral representation \eqref{MIR}. It is worth mentioning that function
$h_N(z)$ is also known explicitly at Ice Point, ($\Delta=1/2$), and Dual
Ice Point, ($\Delta=-1/2$), being expressible in terms of (polynomial) Gauss
hypergeometric function \cite{Ze,CP2}. For instance, at Ice Point the
triple Penner model discussed above generalizes to a two-matrix Penner
model. This model can be studied along the lines presented here, thus
providing a solution to the longstanding problem of limit shape for
Alternating Sign Matrices.

\section*{Acknowledgements}

We thank Nicolai Reshetikhin for useful discussion, and for
giving us a draft of \cite{PR} before completion.
FC is grateful to Percy Deift,  and Courant Institute
of Mathematical Science, for warm hospitality.
AGP thanks INFN, Sezione di Firenze, where part of this work was done.
We acknowledge financial support from MIUR PRIN program (SINTESI
2004). One of us (AGP) is also supported in part by Civilian Research
and Development Foundation (grant RUM1-2622-ST-04), by
Russian Foundation for Basic Research (grant
04-01-00825), and by the program Mathematical Methods in Nonlinear
Dynamics of Russian Academy of Sciences. This work is partially
done within the European Community network EUCLID (HPRN-CT-2002-00325),
and the European Science Foundation program INSTANS.

\end{document}